\documentclass[3p,twocolumn,8pt]{elsarticle}

\usepackage[ruled,vlined]{algorithm2e}
\usepackage{graphicx}
\usepackage{siunitx}
\usepackage{amsmath}
\usepackage{setspace}
\usepackage{natbib}

\newcommand{\I}{\ensuremath{\mathrm{i}}}
\newcommand{\Exp}[1]{\mathrm{e}^{#1}}
\newcommand{\ks}{$k$-space\, }
\newcommand{\norm}[1]{\left\lVert#1\right\rVert}

\journal{Journal of Magnetic Resonance}

\begin{document}

\begin{frontmatter}

\title{Model-driven reconstruction with phase-constrained highly-oversampled MRI}

\author[mymainaddress,mysecondaryaddress]{F. Galve\corref{mycorrespondingauthor}\fnref{eq}}
\ead{fernando.galve@i3m.upv.es}

\author[mymainaddress,mysecondaryaddress]{J. Alonso\fnref{eq}}
\ead{joseba.alonso@i3m.upv.es}

\author[mymainaddress,mysecondaryaddress]{J. M. Algar\'in}
\author[mymainaddress,mysecondaryaddress]{J. M. Benlloch}

\cortext[mycorrespondingauthor]{Corresponding author}
\fntext[eq]{Equally contributing authors}
\address[mymainaddress]{Inst. for Instrum. for Molecular Imaging (i3M), Spanish National Research Council(CSIC), Valencia, Spain.}
\address[mysecondaryaddress]{Universitat Polit\`ecnica de Val\`encia (UPV), Camino de Vera s/n, 46022,Valencia, Spain}

\begin{abstract}
The Nyquist-Shannon theorem states that the information accessible by discrete Fourier protocols saturates when the sampling rate reaches twice the bandwidth of the detected continuous time signal. This maximum rate (the NS-limit) plays a prominent role in Magnetic Resonance Imaging (MRI). Nevertheless, reconstruction methods other than Fourier analysis can extract useful information from data oversampled with respect to the NS-limit, given that relevant prior knowledge is available. Here we present PhasE-Constrained OverSampled MRI (PECOS), a method that exploits data oversampling in combination with prior knowledge of the physical interactions between electromagnetic fields and spins in MRI systems. In PECOS, highly oversampled-in-time $k$-space data are fed into a phase-constrained variant of Kaczmarz's algebraic reconstruction algorithm, where prior knowledge of the expected spin contributions to the signal is codified into an encoding matrix. PECOS can be used for scan acceleration in relevant scenarios by oversampling along frequency-encoded directions, which is innocuous in MRI systems under reasonable conditions. We find situations in which the reconstruction quality can be higher than with NS-limited acquisitions and traditional Fourier reconstruction. Besides, we compare the performance of a variety of encoding pulse sequences as well as image reconstruction protocols, and find that accelerated spiral trajectories in $k$-space combined with algebraic reconstruction techniques are particularly advantageous. The proposed sampling and reconstruction method is able to improve image quality for fully-sampled $k$-space trajectories, while allowing accelerated or undersampled acquisitions without regularization or signal extrapolation to unmeasured regions.
\end{abstract}

\begin{keyword}
Image reconstruction, Magnetic resonance imaging, Iterative reconstruction, Phase constrained methods
\end{keyword}

\end{frontmatter}


\section{Introduction}
The Nyquist-Shannon (NS) theorem specifies that a sampling rate twice the emission bandwidth of a continuous time-dependent signal suffices to recover all the information accessible by Fourier protocols \cite{Landau1967}. This is relevant to Magnetic Resonance Imaging (MRI) because Fourier Transforms (FT) provide an efficient mapping between spatial frequency space (or $k$-space), which is a natural domain of representation for the detected signals, and the sought image space \cite{BkHaacke}. The NS-limit in \ks can be formulated as $\delta k_i = 2\pi/\Delta x_i$, where $\Delta x_i$ is the spatial extent of the Region of Interest (RoI) along the $i$-axis ($i\in \{x,y,z\}$), $\delta k_i$ is the separation between \ks points along $k_i$, and \ks values are given in units of rad/m. From a simple FT perspective, undersampling (i.e. $\delta k_i > 2\pi/\Delta x_i$) produces unwanted aliasing effects in the reconstruction, whereas oversampling ($\delta k_i < 2\pi/\Delta x_i$) is pointless, since there is no useful information to recover at frequencies beyond the NS-limit. However, reconstruction techniques more elaborate than FTs exploit undersampling for scan acceleration \cite{Ravishankar2011}, and oversampling for avoiding aliasing from active spins outside the RoI as well as increasing the signal-to-noise ratio (SNR) and dynamic range of the analog-to-digital converters (ADC) outputs \cite{BkSmith}. We are not aware of the use of oversampling for reconstruction enhancement or scan acceleration in the existing literature (the topics of this work), other than in support-limited extrapolation techniques \cite{Liang1992}, which are unrelated to our approach.

Prior knowledge about the MRI scanner, the sample or the interactions between them can be used to bypass some of the limitations of traditional Fourier methods and NS-limited acquisitions. Examples of prominent techniques exploiting prior knowledge are parallel imaging (PI), compressed sensing (CS) and non-linear gradient (NLG) encoding. In PI, where a phased array of multiple detectors is employed, \ks is typically undersampled along phase-encoded directions \cite{Pruessmann1999}.  Although an FT reconstruction on the signal of any individual detection coil would be aliased, incorporating additional information about the unique spatial sensitivity of every coil in the array allows to recover unaliased images. PI therefore exploits prior knowledge about the scanner. CS, on the other hand, utilizes information about the scanned object. With the advent of CS, it became apparent that the number of required data samples can be related to information content rather than signal bandwidth \cite{Donoho2006}. If the former is sparse in some basis, then fewer samples are necessary, reducing scan times without necessarily sacrificing image quality. Such bases exist, exploiting, for instance, non-local features in the properties of real sampled objects \cite{Wang2001,Lustig2007} or their induced signals in \ks \cite{Luo2018}. Further prior information is the fact that sample spin densities are real quantities, which is at the basis of partial Fourier reconstruction \cite{Haacke1991} and phase-constrained parallel imaging \cite{2017PIreview}. This prior enables sampling only a fraction of $k$-space (half of it if experimental errors are correctly taken into account), reducing acquisition times. Finally, in NLG scanners, where the encoding gradient fields are inhomogeneous and hence spatial frequency and position are not conjugate variables, reconstruction tools other than FTs are a must \cite{Stockmann2010,Cooley2015}. In this case, it is useful to construct an Encoding Matrix (EM) with a priori information about how spin phases are expected to evolve in time depending on their position. The physical model for the interaction between the electromagnetic fields and the sample spins in an MRI system is simple, analytical and highly reliable, so image reconstruction can be performed by inverting the EM and having it act on the discretized signal vector. However, the size of the EM in typical acquisitions is too large for direct inversion and iterative methods are required, so this model-driven approach to reconstruction is used mostly when FTs are not a viable option \cite{Li2013}.

In this paper we consider an MRI method where the acquisition is sampled at rates significantly higher than the Nyquist-Shannon limit and the reconstruction is based on prior knowledge of a) the physical interactions that take place during an MRI scan, and b) the real-valuedness of the spin density of the sample. We call this method PECOS, which stands for PhasE-Constrained OverSampled MRI. In PECOS, abundant data (along frequency-encoded directions) combined with prior knowledge of the interaction model (embodied in an Encoding Matrix) and of real-valuedness of the sample's magnetization (embodied in a projection operation), provides useful information for the reconstruction. We find PECOS a generally valid approach and of added value in some relevant scenarios. In Sec.~\ref{sec:theory} we present the theory behind PECOS. We then benchmark the performance of PECOS with simulations for a variety of encoding pulse sequences and reconstruction methods (Sec.~\ref{sec:results}). We first show that PECOS can perform better than traditional encoding and reconstruction methods for fully sampled \ks trajectories in terms of image quality. We then show that sampling at high rates in Cartesian and Spiral acquisitions can compensate for a reduction in the acquired \ks lines along phase-encoded directions, thereby providing a means of scan acceleration based on prior knowledge distinct from existing approaches such as in parallel imaging or compressed sensing.  In contrast to other methods such as Band-Limited Signal Extrapolation \cite{BLSE} or Low-Rank Modeling of Local \ks Neighborhoods (LORAKS) \cite{Haldar2014}, we do not extrapolate the signal to unmeasured regions. Instead, we fit the highly oversampled data to a unique compatible, phase-constrained, reconstructed image as we explain next. Finally, we include a simple mitigation scheme to incorporate experimental phase errors in the reconstruction algorithm that works well for Cartesian sequences.

\section{Methods}
\label{sec:theory}

\subsection{Prior knowledge for spectral reconstruction - an example}

\begin{figure}[!h]
	\centering
	\includegraphics[width=\columnwidth]{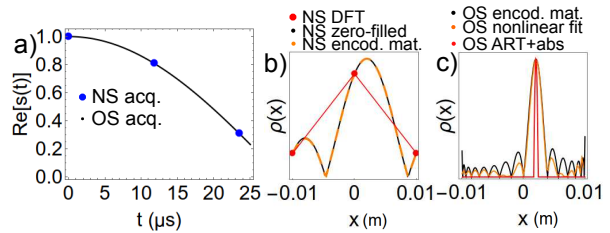}
	\caption{Signal and reconstruction of a single spin sample, placed at $x_0=2$~mm, inside an RoI going from -10 to 10~mm. a) Normalized signal acquired according to the NS-limit ($\delta t=\SI{12}{\micro s}$, large blue dots) and oversampled signal ($\delta t=\SI{0.12}{\micro s}$, small black dots), with $t_{\text{acq}} = \SI{25}{\micro s}$, and $g=100$~mT/m. b) Image reconstructed by FT from NS-sampled signal: the red points correspond to the DFT; the black curve has been zero-filled to high $k$ values for a quasi-continuous reconstruction; and the dashed orange line is the reconstruction by EM from the same NS-sampled signal. c) Reconstruction by EM (black), Mathematica's NonlinearModelFit function (orange), and ART with absolute value projection (red, $\lambda=0.1$, $10^5$ iterations), from the oversampled signal. For the first two, the resolution starts to saturate below $\delta t \sim \SI{1}{\micro s}$ while for projected ART resolutions increases further. To obtain the OS-EM resolution in c), an FT reconstruction requires $t_{\text{acq}} = \SI{100}{\micro s}$. }
	\label{fig:CONCEPT}
\end{figure}

It has been long known that fitting measured data to a presumed image shape may outperform FT-based reconstruction in MRI. For instance, fitting a collection of boxcar functions eliminates Gibbs ringing and enables superresolution \cite{Haacke1989}. This line of reasoning developed into the field of linear prediction \cite{Makhoul1975}, with specific methods such as LORAKS \cite{Haldar2014}. The main insight behind this performance enhancement is that the Nyquist-Shannon sampling limit underlying Discrete Fourier Transform (DFT) methods is agnostic of any prior knowledge of the physical acquisition model: it simply takes equispaced \ks data as an input, and performs an orthogonal transformation. However, the time evolution of spins during an MRI scan is known a priori, and fitting the acquired signal using an interaction model can result in higher fidelity reconstructions.\\
\indent For illustration purposes, let us consider a single spin (located at $x_0$) and a constant 1D magnetic gradient of strength $g$ (in T/m). After excitation, the detected signal is expected to evolve as $\Exp{-\I \gamma g x_0 t}$, with $\gamma$ the gyromagnetic factor ($\approx 2\pi\cdot\SI{42}{MHz/T}$ for protons). FT reconstruction requires $\delta k= 2\pi/\Delta x = \gamma g \delta t$ (with $\delta t$ the dwell time of the ADC) for a time $t_{\text{acq}}=k_{\text{max}}/(\gamma g) =2\pi/(\gamma g\delta x)$ to determine the position of the spin with a resolution $\delta x$. Increasing the sampling rate above the NS-limit will not improve the reconstruction, whereas a simple fit to the known model can yield better results. This is illustrated in Fig.~\ref{fig:CONCEPT}, where we follow five different protocols: (1) FT reconstruction of an NS-limited acquisition, (2) fit with an EM ($M$) that relates the signal $\vec{s}=\{s(t=i \delta t)\}$, $i\in [0,t_{\text{acq}}/\delta t] $ and the image $\vec{\rho}=\{\rho(x=-\Delta x/2+j\cdot \delta x)\}$, $j\in [0,\Delta x/\delta x ] $, such that $\vec{s}=M\vec{\rho}$, and invert $M$ to obtain $\vec{\rho}=M^{-1}\vec{s}$, (3) idem but oversampling with respect to the NS-limit, (4) nonlinear fitting given by Mathematica with oversampling and (5) our proposed algorithm: Kaczmarz's algebraic reconstruction technique (ART) \cite{Kaczmarz1937,Gower2015} with projection to absolute values from the oversampled signal (see next section). Methods (3-4) clearly outperform the reconstruction capabilities of FT methods (Fig.~\ref{fig:CONCEPT}c) through the use of oversampling, while adding projection to absolute values (5) yields further performance enhancement. The latter projects $\vec{\rho}$ to its absolute value, since we know that, without experimental errors, our sample has a real spin density. This nonlinear operation incorporates prior phase knowledge to the reconstruction, forcing phases to be those given by pure gradient encoding.\\
\indent This simplified example does not fully capture the complexity of real-life MRI signals and procedures, but it does show how prior knowledge can be exploited. In fact we have checked that when more point-like spins are included in the sample, the resolution capabilities diminish: there is a trade-off between sample complexity and the final ability of the algorithm to fit the signal to a reconstructed image. In Sec.~\ref{sec:results} we show the performance of PECOS in more realistic (even if simulated) scenarios.

\subsection{Interaction model and reconstruction in PECOS}

In model-driven reconstruction, it is common to express the acquired discrete signal $\{s(t_i)\}$ as a vector $\vec{s}$ of length equal to the number of time steps $N_t$, and the sought spin density as a vector $\vec{\rho}$ of length equal to the number of voxels $N_r$. The sensor ($\vec{s}$) and image ($\vec{\rho}$) domains are related by $M(\vec{r},t)=\Exp{-\I\gamma \vec{k}(t)\cdot\vec{r}}$ as $\vec{s} = M\vec{\rho}$, where $\vec{k}(t)$ is the time integral of the applied gradient up to time $t$. Thus, $M$ is the Encoding Matrix that stems from the physical interaction model and acts as a prior in PECOS and other model-driven methods (such as NLG encoding). For instance, for a 2D NS-limited acquisition with a sought resolution ($\delta x,\delta y$) and size of the RoI ($\Delta x,\Delta y$), where $x$ ($y$) is a frequency (phase) encoded direction, the EM size is equal to the square of the number of pixels, where $N_r = N_x N_y = (\Delta x/\delta x)(\Delta y/\delta y)$, the first EM row corresponds to $t=0$ and the last one to $t_\text{acq} = 2\pi N_y/(\gamma G_x \delta x)$, in increments of $\delta t =2\pi/(\gamma G_x \Delta x)$.

One can solve for $\vec{\rho}$ by direct inversion of the Encoding Matrix as $\vec{\rho}=M^{-1}\vec{s}$, or by any other means of solving the system of linear equations, e.g. by iterative algorithms such as algebraic reconstruction techniques (ART, \cite{Gordon1970}) or conjugate gradient methods \cite{Fletcher1964}. In order to avoid the linear problem from being under-determined, $N_t \geq N_r$ is required (for DFT protocols $N_t$ and $N_r$ must be equal). Thus, $M$ has dimension at least $N_t\times N_r\geq N_r^2$, and scales as $\text{dim}(M)\geq N_x^{2n}$ for an $n$-dimensional acquisition with $N_x$ pixels per dimension. Matrix inversion is thus not a scalable approach, and can lead to strong noise amplification if the EM's condition number is high. Instead, most of the below reconstructions result from running the Kaczmarz method in a 16-thread AMD Ryzer processor, or in a Graphics Processing Unit with Cuda \cite{BkSanders} for bigger samples. This method is a phase constrained modality of ART which performs the steps given in Algorithm~\ref{alg1}, where $\lambda$ is the update parameter, $^*$ denotes complex conjugation, and $\vec{M}_t$ is the vector formed by the $t$-th row of the Encoding Matrix $M$, with components $(\vec{M}_t)_r:=M_{t,r}$.
\begin{algorithm}
\SetAlgoLined
 $\bullet$ initialization: set $\vec{\rho}=0$\\
 $\bullet$ reconstruction:\\
 \For{$n\in [1,N_{\text{its}}]$}{
            \For{$t\in [1,N_t]$}{
         \hspace{4mm} $\vec{\rho}\leftarrow\vec{\rho}+\lambda\frac{s_t-\vec{M}_t\cdot\vec{\rho}}{|\vec{M}_t|^2} \vec{M}_t^*$\\
         \hspace{4mm} $\vec{\rho}\leftarrow|\vec{\rho}|$}}
 \caption{Phase-Constrained ART}
 \label{alg1}
\end{algorithm}

\begin{figure}[!h]
	\centering
	\includegraphics[width=0.8\columnwidth]{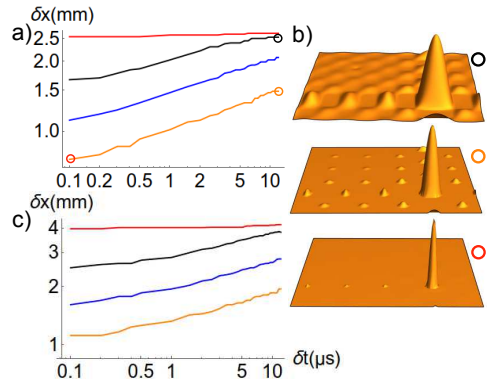}
	\caption{2D reconstruction of a single spin sample, placed at $x_0=y_0=5$~mm, inside an RoI going from -10 to 10~mm. a) FWHM of reconstructed image by a 2D EPI sequence vs $\delta t$
 	($\delta t=\SI{12}{\micro s}$ amounts to Nyquist sampling), with $t_{\text{acq.}} = 1$~ms, and $g=100$~mT/m; ART parameters $\lambda=$0.1, 10 (black), 100 (blue), 1000 (orange) iterations with projection to absolute values, and 10 (red) iterations without projection. b) 2D reconstructed images for the parameter points highlighted in a). c) Idem, for a spiral acquisition.
}
	\label{fig:CONCEPT2D}
\end{figure}

\indent  The method, hence, runs over all time steps, and we iterate $N_\text{its}$ times, so that the total updates to $\vec{\rho}$ are $N_t\times N_\text{its}$. Since ART is merely an $l_2$-norm minimization  method, one can include also terms penalizing the update step \cite{Li2013,Liu2011}. We discuss this possibility in Sec.~\ref{sec:penalties}. 

In figure~\ref{fig:CONCEPT2D} we revisit the case of a single spin, now in 2D, following an Echo Planar Imaging sequence (EPI, \cite{EPI}). There we plot the full width at half maximum (FWHM) of the reconstructed image peak, the point-spread-function (PSF), as a function of oversampling and ART iterations. We see that without phase-projection (red) there is no improvement in peak width nor reduction of sidelobes. In contrast, the phase projection in Algorithm~\ref{alg1} combined with oversampled acquisition beyond NS, is able to progressively improve both. This behavior is at the basis of the improvements reported in this work.

\section{Results}
\label{sec:results}

In the remainder of this paper we use the following minimal settings unless otherwise explicitly stated: single-coil reception, no regularization (total variation, or others), and a direct comparison between fully-sampled FT with an oversampled-in-time phase-projected ART (see Alg.~\ref{alg1}). All simulated signals are generated from a phantom of higher resolution than the reconstructions, chosen so that the number of pixels in the former is non-divisible by the pixels in the latter. Besides, pixels are considered dense spin distributions rather than a single ``heavy'' spin. These distributions are integrated to find the contribution to the overall signal from every individual pixel. Repeatability of results has been checked against a $\times 4^2$ increase in the number of dense pixels. We take these precautions to avoid ``inverse crime'' situations \cite{Guerquin-Kern2012}.

\subsection{PECOS with fully-sampled \ks trajectories}
\label{sec:fullysampled}
\begin{figure}
	\centering
	\includegraphics[width=\columnwidth]{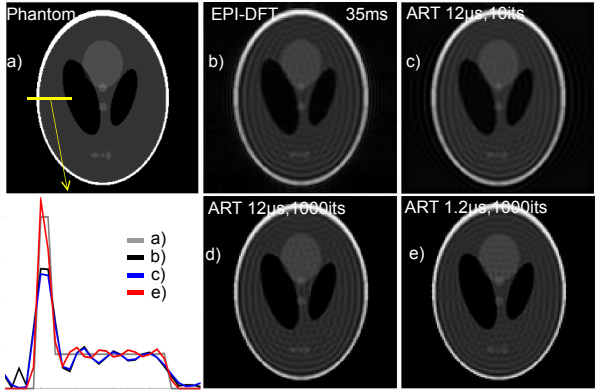}
	\caption{Shepp-Logan phantom reconstructions ($\Delta x=\Delta y=2$~cm) with Cartesian \ks NS. Here, $t_\text{acq}=35$~ms at a gradient strength $g=100$~mT/m. The phantom (a) is made of 256x256 dense pixels and has been reconstructed by: DFT and NS sampling (i.e. $\delta t=\SI{12}{\micro s}$, b); 10 iterations of ART with phase projection and NS sampling (c); idem, with 100 iterations (d); and $\times 10$ oversampling and 1000 ART iterations (e). We plot the 1D profile shown in yellow in a) for each of the reconstructions, except d), showing the improvement in Gibbs ringing and reconstruction accuracy.}
	\label{fig:SHEPP}
\end{figure}

\begin{figure}[h!]
	\centering
	\includegraphics[width=0.9\columnwidth]{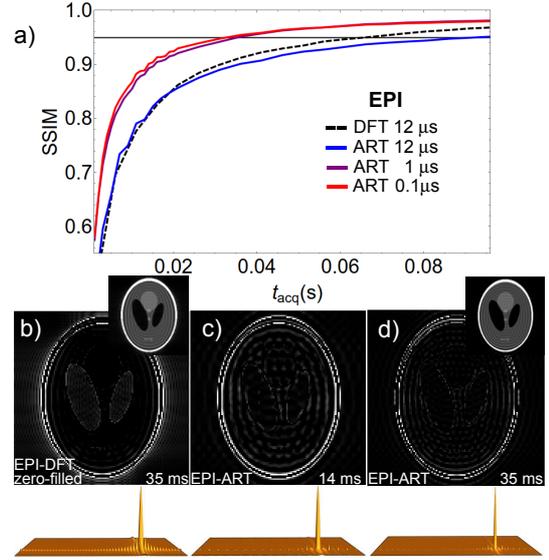}
	\caption{a) Reconstructed image quality (SSIM) as a function of acquisition time. Black-dashed: NS-limited EPI with DFT ($\delta t=\SI{12}{\micro s}$, zero-filled to reach 120x120 pixels). Continuous lines: EPI with ART at $\delta t=\SI{12}{\micro s}$ (blue), $\times 12$ oversampling ($\SI{1}{\micro s}$, purple), and $\times 120$ oversampling ($\SI{0.1}{\micro s}$, red). b) Reconstruction difference with the original phantom, for NS-limited DFT at $t_{\text{acq}}=\SI{35}{ms}$. Total absolute error $\approx 4~\%$. c) $\times 120$ oversampling and ART at $t_{\text{acq}}=\SI{14}{ms}$, $\delta t=\SI{0.1}{\micro s}$. Total absolute error $\approx 3.8~\%$. d) $\times 120$ oversampling and ART at $t_{\text{acq}}=\SI{35}{ms}$, $\delta t=\SI{0.1}{\micro s}$. Total absolute error $\approx 2.5~\%$. For all plots $g=100$~mT/m, $\text{RoI}=(2$~cm)$^2$ and reconstruction on 120x120 pixels. The insets show the NS-limited DFT and oversampled ART reconstructions. The PSFs for each sequence and reconstruction method are shown at the bottom.}
	\label{fig:EPIfull}
\end{figure}
We find PECOS can perform better than standard methods for fully sampled (non-accelerated) Cartesian trajectories, consistent with sidelobe reduction in the PSF.
To check the behavior observed in Fig.~\ref{fig:CONCEPT2D} with a sample with more complexity, we choose a Shepp-Logan phantom. In  Fig.~\ref{fig:SHEPP} we compare DFT reconstruction with NS sampling against our algorithm with different levels of oversampling and ART iterations. While the result does not lead to superresolution as observed in Fig.~\ref{fig:CONCEPT2D}, there is an increase in definition of the inner structures in the phantom. In both cases we see a better approximation to the original shape, while Gibbs ringing gets progressively softened.

To better quantify these differences we employ the structural similarity index (SSIM, \cite{SSIM}). In Fig.~\ref{fig:EPIfull}a), the dashed black line shows the SSIM of a Shepp-Logan phantom imaged with a single-shot EPI sequence and reconstructed with DFT, as a function of the readout duration $t_\text{acq}$. An ART reconstruction from the same data shows a slightly worse behavior with $\lambda = 0.1$ and $N_\text{its} =10$ (solid blue). When we oversample in time by a factor of 12 ($\delta t = \SI{1}{\micro s}$), ART reconstructions result in a higher SSIM (solid purple). For reference, the DFT line crosses an SSIM of 0.9 at $t_\text{acq} \approx \SI{35}{ms}$, whereas the ART line for $\delta t = \SI{1}{\micro s}$ reaches the same value at $\approx \SI{14}{ms}$. Further oversampling does not significantly improve reconstruction quality with these parameters (solid red line).

Images b)-d) in Fig.~\ref{fig:EPIfull} correspond to the pixel-by-pixel difference between the phantom and reconstructed images. The SSIM for the left (DFT) and middle (ART \SI{0.1}{\micro s}) images are both $\approx 0.9$, and their total absolute errors (sum of normalized absolute value deviations in pixel brightness) are both at the 4~\% level, although the middle plot was acquired in less than half the time (14 vs 35~ms). A 35~ms oversampled acquisition can be reconstructed with ART with an $\text{SSIM}\approx 0.95$ and a total absolute error $\approx 2.5$~\%. The PSFs indicate that oversampling at 14~ms results in reduced sidelobes and a similar spatial resolution to DFT at 35~ms. Although we have used SSIM and absolute value deviation errors, similar behavior is seen for other quantifiers such as RMS.

\subsection{Imaging with reduced \ks coverage}
\label{sec:under}

\begin{figure}
	\centering
	\includegraphics[width=0.8\columnwidth]{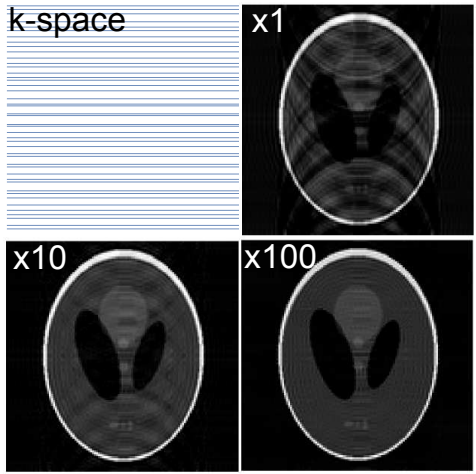}
	\caption{Shepp-Logan phantom reconstructions ($\Delta x=\Delta y=2$~cm) with Cartesian \ks sampling and x2 acceleration in $k_y$. Here, $t_\text{acq}=50$~ms at a gradient strength $g=100$~mT/m, so $k_{\text{max}}=14500$~rad/m. The images show ART reconstructions with oversampling factors of $1,10,100$ along $k_x$. ART reconstruction parameters: $\lambda=0.1$, $N_\text{its}=40$. The scanned \ks lines (top left plot) have been randomly displaced by up to 10~\% of the NS separation to suppress aliasing.}
	\label{fig:fig1}
\end{figure}

We next check the performance of PECOS with standard Cartesian sampling at the NS rate, but removing every second phase-encoded \ks line (Fig.~\ref{fig:fig1}). In the language of Parallel Imaging this corresponds to a two-fold undersampling or acceleration, and requires the use of two signal receiver coils with complementary sensitivity regions. In the case of partial Fourier reconstruction, only half the \ks is needed, so we expect that our real-projected iterative algorithm can also reduce by half the required $k$-space lines. The images in Fig.~\ref{fig:fig1} show PECOS reconstructions assuming a single detector: these show that sampling at high rates in a Cartesian acquisition and projecting $\rho$ to absolute values can compensate for a reduction in the acquired \ks lines along a phase-encoded direction, thereby accelerating the acquisition with a model-driven approach. The reconstruction quality increases with oversampling in time, i.e. along the frequency-encoded direction, although at $\times100$ oversampling we still see some small remaining artifacts. These acquisitions are simulated with small random displacements of the scanned \ks lines which suppress aliasing artifacts. In a different set of simulations, we have also successfully reconstructed images from acquisitions with variable density sampling, where the central region of \ks is NS-limited along the phase-encoded directions and the outer portions of \ks are undersampled. This is typical in scans with reconstructions based on compressed sensing \cite{Lustig2007}.

\begin{figure}
	\centering
	\includegraphics[width=0.8\columnwidth]{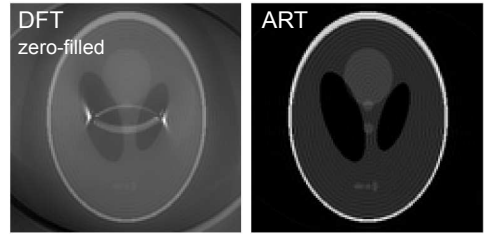}
	\caption{\footnotesize Shepp-Logan phantom reconstruction with spiral pulse sequence accelerated by $\times 2$ with $t_{\text{acq}}=50$~ms. Left: DFT reconstruction with $\delta t= \SI{12}{\micro s}$. Right: ART reconstruction with $\delta t= \SI{0.12}{\micro s}$, $\lambda=0.1$, $N_\text{its} = 10$.}
	\label{fig:Spix2}
\end{figure}

PECOS can also be used to reconstruct images from spiral sequences with radial undersampling (accelerated by a factor $\alpha$), i.e. where instead of $k_x(t)=\theta(t)\cos \theta(t)$ (and sine for $k_y$) with a slew-rate $S$, $k_x (t)=\alpha\theta(t)\cos \theta(t)$ and $S/ \alpha$ is used, to keep the effective slew-rate at the same value \cite{spiralGlover}. In Fig.~\ref{fig:Spix2} we compare reconstruction of a x2 accelerated spiral acquisition by DFT (regridded into a Cartesian Nyquist $k$-space) and phase-constrained oversampled ART algorithm: not only does ART enable x2 undersampled acquisition, but it also avoids typical circular artifacts appearing for spiral regridded trajectories. Importantly, for spiral acquisitions the use of oversampling increases the number of \ks points, i.e. the amount of plane waves with different propagation directions that constitute the description of the sample image, thus helping remove aliasing caused by the x2 acceleration. Therefore, in non-Cartesian acquisitions the reconstruction improvements given by our protocol may become particularly relevant.\\
\indent Doubling the \ks step with Fourier protocols is equivalent to halving the RoI, so aliasing artifacts are expected, as seen in Fig.~\ref{fig:Spix2}. For the left image, we recast the data into a Cartesian grid (regridding) to use DFT protocols, but this does not prevent aliasing due to the undersampling in the radial \ks direction. In contrast, the ART reconstruction appears unaliased when the signal is sufficiently oversampled in time. For the case of x2 accelerated spiral acquisition, NS sampling with high number of ART iterations is strictly worse than an oversampled acquisition, highlighting the importance of the extra non-NS \ks points to avoid aliasing effects.

\subsection{Performance under phase errors}
\label{sec:phase}
\begin{figure}[!h]
	\centering
	\includegraphics[width=7cm]{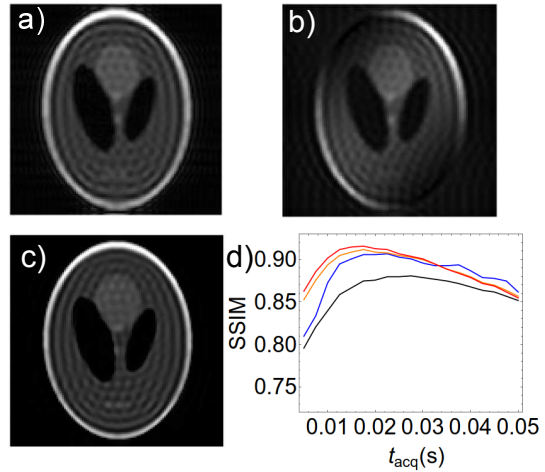}
	\caption{\footnotesize Single-shot EPI reconstruction, with $t_{\text{acq}}=16$~ms, with residual gradients yielding $\SI{2}{\micro T}$ deviation at edges of RoI (2~cm size). a) DFT, b) ART without phase mask ($\delta t=\SI{1.2}{\micro s}$), c) ART with phase mask ($\delta t=\SI{1.2}{\micro s}$). d) We show SSIM of the reconstruction vs acquisition time when residual gradients yield $\SI{1.2}{\micro T}$ deviation at the edges of the RoI, for DFT(black), ART with $\delta t=\SI{12}{}$, $\SI{6}{}$, $\SI{3}{\micro s}$ (blue, orange, red). ART parameters: $\lambda=$~0.1, 100 iterations.
	}
	\label{fig:PHASE1}
\end{figure}

Even after careful calibration of experimental setups, phase errors are inevitably present due to field inhomogeneities, eddy currents, magnetic susceptibility changes in tissue borders or chemical shift, and they lead to artifacts in the reconstructed image \cite{phaseErrors}. These errors, caused by an incorrectly predicted phase evolution, are more prominent in single-shot sequences such as spiral and EPI and can be mitigated by splitting the $k$-space acquisition into several excite/acquire blocks. While spiral and radial sequences with phase errors lead to blurring, in Cartesian sequences they can be accounted for by an image phase map, i.e. the reconstructed image is $\rho_{\text{exp}}=\exp[i\phi(\vec{r})]\rho_{\text{real}}$, where the phase map $\phi(\vec{r})$ typically depends on the particular parameters of the sequence. Since our reconstruction protocol partially hinges on the projection of image solutions into the space of real images, phase errors are of critical importance. One possible strategy is to incorporate previously calibrated errors into the encoding matrix. A preferable alternative is to use data from the single-shot sequence itself to account for phase errors. With the latter approach we are able to show that our reconstruction protocol still improves image quality with respect to DFT, as can be seen in Fig.~\ref{fig:PHASE1}. With a single-shot EPI sequence with residual gradients of $\SI{2}{\micro T}$ at the edges of the RoI, we use the Nyquist grid $k$-space points to estimate a phase map through a DFT, and incorporate this phase map into the ART algorithm. Considering that our acquired signal is $\vec{s}=M\vec{\rho}_{\text{exp}}$ with $M$ the ideal encoding matrix without phase errors, and $\vec{\rho}_{\text{exp}}$ the resulting complex image, we can incorporate the phase map into the encoding matrix as $\vec{s}=M\vec{\rho}_{\text{exp}}=M e^{i\phi(\vec{r})}\rho_{\text{real}}=\tilde{M}\rho_{\text{real}}$, and then use the same ART algorithm that projects $\rho$ in the space of real solutions. We show several reconstructions with $\SI{2}{\micro T}$ to exaggerate the effect, but use $\SI{1.2}{\micro T}$ in Fig.~\ref{fig:PHASE1}d to calculate SSIM vs acquisition time, corresponding to $100$~Hz frequency span due to residual gradients. We have obtained the phase map from a DFT of the signal up to $k_{\text{max}}=1000$~rad/m, which gives a low-resolution estimation of the map, but DFT of the full acquisition could be used for situations where high-resolution phase errors are present, such as due to susceptibility changes in tissues borders. In the simulated case, $\phi(\vec{r})$ does not wrap in the RoI, otherwise we would need a slightly more elaborate procedure.
\begin{figure}[h!]
	\centering
	\includegraphics[width=\columnwidth]{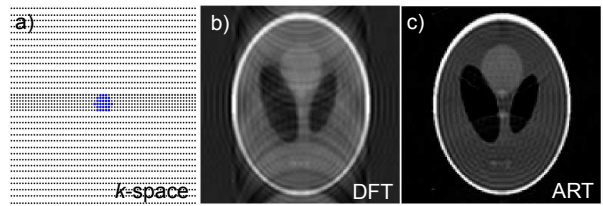}
	\caption{\footnotesize Multi-shot partially undersampled reconstruction, with $k_{\text{max}}=10300$~rad/m, with residual gradients yielding $\SI{1.2}{\micro T}$ deviation at edges of RoI (2~cm size). a) A central disk (a hundredth of the total $k$-space) of radius $k_{\text{max}}/10$ , highlighted with blue points, is scanned according to Nyquist spacings, while the rest has been undersampled by removing half of horizontal lines (x2 undersampling). b) DFT reconstruction. c) ART reconstruction with phase correction with $\lambda=$~1, 100 iterations.}
	\label{fig:PHASE2}
\end{figure}

We have also checked that a partially undersampled multi-shot Cartesian acquisition can be corrected, where we use a central Nyquist-scanned portion of $k$-space to estimate the low-resolution phase map to later correct the encoding matrix to be used by ART. The rest of \ks is x2 undersampled with a slight vertical random shift of lines to suppress aliasing. The result can be seen in Fig.~\ref{fig:PHASE2}. In multi-shot sequences, phase errors are reset to zero at each shot, i.e. for each $k$-space line, and thus are less severe. Finally, in the case of non-Cartesian acquisitions, more elaborate strategies can mitigate phase errors by estimating accumulated phase errors from two consecutive, time-delayed, acquisitions, as in \cite{ARTphase}, where an ART algorithm with a phase estimation step accounts for phase error accrual.

\subsection{Penalties}
\label{sec:penalties}
The ART algorithm (Algorithm~\ref{alg1}) is a gradient descent method where the data consistency condition $\vec{s}=M\vec{\rho}$ is enforced. Consider the $l_2$-norm squared cost function $\norm{\vec{s}-M\vec{\rho}}^2$; every ART step implements its gradient at a given time, and advances along $\vec{M}^*_t$ the reconstructed image $\vec{\rho}^{(n)}$ by an amount given by the data consistency error at that time step $s_t-\vec{M}_t\cdot \vec{\rho}^{(n)}$. Since $M$ is a Vandermonde matrix, its rows $\vec{M}_t$ are linearly independent vectors, and thus span a whole set of directions \cite{Koehl1999}.  Similarly, other $l_2$-norm penalties, e.g. Tikhonov regularization with $\norm{\vec{\rho}}^2$, may be included by incorporating its derivative into the ART steps (i.e. adding a term $\beta \vec{\rho}$, where $\beta$ is the new update parameter, see Ref.~\cite{Li2013}). 

\begin{figure}[!h]
	\centering
	\includegraphics[width=\columnwidth]{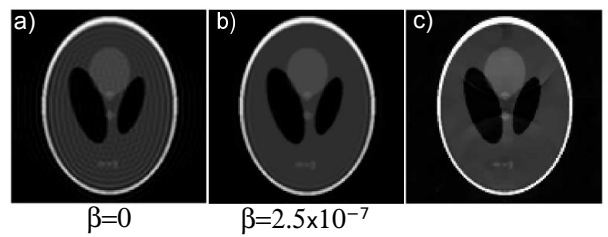}
 	\caption{\footnotesize a) Shepp-Logan phantom reconstruction with an oversampled EPI pulse sequence with $t_{\text{acq}}=50$~ms, $\delta t=\SI{3}{\micro s}$, $\lambda=0.1$, $N_{\text{its}}=10$. b) The same, adding a total variation penalty to the ART algorithm (see text). c) Figure~\ref{fig:PHASE2}c (multi-shot, undersampled acquisition) with penalty added ($\beta=10^{-6}$).}
	\label{fig:TV}
\end{figure}

Adding $l_1$-norm penalties by the gradient method is ill-defined because they imply expressions of the form $\norm{x}_1=\sum_i |x_i|$, whose derivative has singularities. One possibility is to replace the gradient operator by a proximal operator as in compressed sensing. Another is to add a small $\epsilon$ term ($|x_i| \simeq \sqrt{x_i^2+\epsilon}$) and build it into the ART step. An example of such procedure is the total variation (TV), which quantifies the total spatial derivative of $\vec{\rho}$. The $l_1$-norm of total variation can be expressed  as
$$\norm{\vec{\rho}}_{_\text{TV}} =\sum_{i,j}\sqrt{(\rho_{i+1,j}-\rho_{i,j})^2+(\rho_{i,j+1}-\rho_{i,j})^2+\epsilon}$$
and its gradient can be incorporated into the ART algorithm \cite{TV,TV2} as given in Algorithm~\ref{alg2}.
\begin{algorithm}
\SetAlgoLined
 $\bullet$ initialization: set $\vec{\rho}=0$\\
 $\bullet$ reconstruction:\\
 \For{$n\in [1,N_{\text{its}}]$}{
            \For{$t\in [1,N_t]$}{
         \hspace{4mm} $\vec{\rho}\leftarrow\vec{\rho}+\lambda\frac{s_t-\vec{M}_t\cdot\vec{\rho}}{|\vec{M}_t|^2} \vec{M}_t^*+\beta \nabla \norm{\vec{\rho}}_{_\text{TV}}$\\
         \hspace{4mm} $\vec{\rho}\leftarrow|\vec{\rho}|$}}
 \caption{Phase Constrained TV-ART}
 \label{alg2}
\end{algorithm}
Adding TV regularization penalizes stark brightness differences among neighboring pixels, which for a $l_2$-norm smoothens (low-pass filters) the reconstructed image. However, the $l_1$-norm tries to minimize the amount of pixels where such strong brightness changes occur. Figure~\ref{fig:TV} shows an example where Gibbs ringing effects are alleviated as we increase $\beta$, improving image quality without smoothing effects. We also revisit the undersampled EPI case with phase errors in Fig.~\ref{fig:TV}c to show that TV penalty can also improve image quality in such unfavorable situations.

An open research direction is to explore the combination of ART with oversampling and rigorous $l_1$-norm regularization. This could be relevant in the context of CS, where an $l_1$-norm penalty is imposed on e.g. the wavelet-basis coefficients of the reconstructed image, which are typically few because natural images have a sparse representation in that basis. In order to incorporate $l_1$-norm penalties without the $\epsilon$-regularization of the $l_1$-norm, one may use a generalized concept of gradient (the proximal operator), leading to more elaborate methods such as Alternating Direction Method of Multipliers (ADMM \cite{ADMM}) or Split-Bregman \cite{splitBregman} protocols.

\subsection{ART parameters}

\begin{figure}[!h]
	\centering
	\includegraphics[width=0.7\columnwidth]{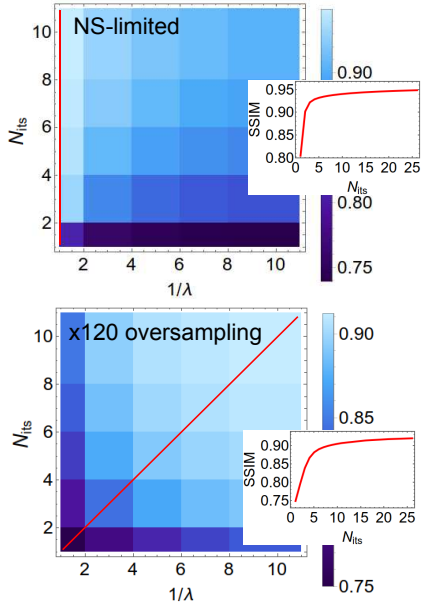}
	\caption{\footnotesize SSIM of Shepp-Logan reconstructions at 120x120 pixels vs $\lambda$ and $N_{\text{its}}$. Top: NS-limited acquisition ($\delta t=\SI{12}{\micro s}$), with $t_\text{acq}=20$~ms. The inset shows the SSIM along $\lambda=1$ line. Bottom: x120 oversampled acquisition ($\delta t = \SI{0.1}{\micro s}$), with $t_\text{acq}=10$~ms. The inset shows the SSIM along the $\lambda\times N_{\text{its}}=1$ line.}
	\label{fig:ARTparam}
\end{figure}

We explore here the influence of ART parameters ($\lambda$, $N_\text{its}$) on the quality of reconstruction for a fully-sampled EPI sequence, by comparing the SSIM of ART outputs to an original Shepp-Logan phantom (see Fig.~\ref{fig:ARTparam}). As a rule of thumb, we find satisfactory results when $\lambda\times N_\text{its} \approx 1$ for oversampled acquisition, in agreement with the bottom plot. For NS-limited acquisitions ($\delta t=\SI{12}{\micro s}$ in Fig.~\ref{fig:ARTparam}) it is often better to use $\lambda=1$ and iterate extensively, even though artifacts may appear. In general~\cite{Algarin2020}, we find that about 10 iterations is an acceptable compromise between computation time and reconstruction finesse. For special cases with undersampling, phase errors or when using penalties, it is often convenient to explore further iterations.

\begin{figure}[!h]
	\centering
	\includegraphics[width=0.75\columnwidth]{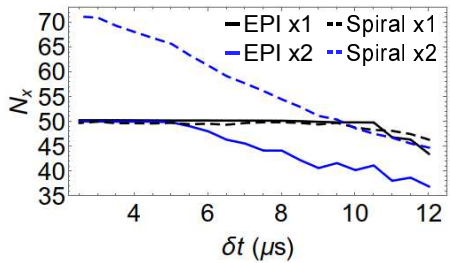}
	\caption{\footnotesize Number of significant pixels that can be obtained from the encoding matrix, calculated as the square root of the number of singular values $\geq 10$~\% of the maximum singular value, plotted as a function of $\delta t$. Continuous lines: EPI with $t_\text{acq}=30$~ms. Dashed lines: spiral with $t_\text{acq}=30$~ms. Acceleration factors are $\times 1$ (black) and $\times 2$ (blue). Here, $g=100$~mT/m, $\Delta x=\Delta y=2$~cm.}
	\label{fig:SINGULAR}
\end{figure}

\section{Discussion}
\label{sec:discu}

In an attempt to find a quantitative explanation to the potential of PECOS, we have checked that the rank of the EM, which gives the number of linearly independent rows, loosely determines part of its resolution power and limitations. The rank corresponds to the number of individually resolvable pixels, i.e. those whose contribution to the detected signal is orthogonal to all of the rest's. Figure~\ref{fig:SINGULAR} shows $N_x$ (the square root of the EM rank) corresponding to Cartesian (EPI) and spiral acquisitions similar to those used for results in this paper. Here we see the EM rank predicts that the accelerated spiral performs better than NS-limited EPI and spiral acquisitions, as is expected by the fact that it can reach a higher $k_{\text{max}}$. Also, a two-fold accelerated Cartesian acquisition can perform as well as a fully-sampled trajectory if the former is sufficiently oversampled along the frequency-encoded direction (in Fig.~\ref{fig:fig1} some artifacts remain for OSx100, but we checked that more ART iterations can neutralize them). Similarly, we see that the advantage due to oversampling seems to saturate for short enough dwell times, consistent with the fact that rows differ by negligible amounts for sufficiently small time steps (see Fig.~\ref{fig:EPIfull}). Nevertheless, the EM rank on its own does not suffice to quantify the expected performance of a given sequence, as it lacks a requirement on \ks coverage. For instance in undersampled sequences, even if the rank predicts that one can reconstruct as well as if fully-sampled, projection of $\vec{\rho}$ to real values is necessary to remove aliasing. Furthermore, EPIx2 does not reach higher $N_x$ than EPIx1 in Fig.~\ref{fig:SINGULAR}, even if it can reach a higher $k_{\text{max}}$ in the same scan time. This seems to point at the rank being partially a measure of potential reconstruction capability, although it does not fully capture the requirements for obtaining a non-aliased image from the undersampled signal.

Looking ahead, there are relevant open avenues besides how to quantify the goodness of a given encoding scheme. One particular aspect is to extend these studies to sequences with a more complex structure of resonant radio-frequency pulses than we have assumed here (see, e.g., Ref.~\cite{Bieri2013}), or with simultaneous radio-frequency excitation and detection \cite{Sohn2016}. Also, the priors exploited for scan acceleration by PECOS, parallel imaging and compressed sensing are of different nature, suggesting that a combination should be possible for enhanced performance with respect to the results we have presented. Finally, all these ideas need to be experimentally validated. A critical requirement in this sense is the direct access to raw, unfiltered data from the readout electronics. This is not necessarily a given in many MRI laboratories.

\section{Conclusion}
\label{sec:concl}
In this paper we have presented PECOS, an encoding and reconstruction method combining data sampling at rates well above the Nyquist-Shannon limit, with phase-constrained algebraic reconstruction techniques. We have demonstrated unaliased reconstructions from accelerated acquisitions (with reduced \ks coverage) and higher quality images from fully sampled \ks trajectories than are possible with Nyquist-Shannon-limited acquisitions and Fourier-based reconstruction. Phase constraint seems to enable reconstruction of undersampled trajectories, as well as sidelobe reductions in the PSF, while oversampling seems to allow for better image fitting to the prior physical model, concretized in an Encoding Matrix. The implementation of the algorithm is straight-forward and can easily be extended to approximate $l_1$-norm penalties. For some simple experimental errors, such as unshimmed gradients in Cartesian trajectories, phase error correction can be easily included without further acquisitions, even in the case of a x2 undersampled scan.

\section{Declaration of competing interests}
The authors declared that there is no conflict of interest.

\section{Data availability}
We have made available the codes used in this paper at github~\cite{git}.

\section{Contributions}
{\bf Fernando Galve}: Conceptualization, Investigation, Simulations, Data Analysis, Writing- Original draft preparation. {\bf Joseba Alonso}: Conceptualization, Investigation, Data Analysis, Writing- Original draft preparation. {\bf Jos\'e Miguel Algar\'in}: Data Analysis, Writing- Reviewing and Editing. {\bf Jos\'e Mar\'ia Benlloch}: Conceptualization, Writing- Reviewing and Editing.

\section{Acknowledgements}
We thank Andrew Webb (LUMC), Peter Börnert (LUMC, Philips) and Justin Haldar (USC) for discussions. We thank Jos\'e Miguel Alonso (GRyCAP-UPV) and Ignacio Blanquer (GRyCAP-UPV) for help on CUDA programming. This work was supported by the European Commission under Grant 737180 (FET-OPEN: HISTO-MRI) and the Spanish Ministry of Science, Innovation and Universities (MICINN) through program “Proyectos I+D+i 2019” (PID2019-111436RB-C21).

%

\end{document}